\documentclass[12pt]{iopart}

\usepackage{epsfig}
\usepackage{graphicx}
\usepackage{dcolumn}
\usepackage{bm}

\begin{document}

\title{Nitrogen and fluorine doped ZrO$_2$: A promising $p$-$n$
junction for ultraviolet light-emitting diode}
\author{Sudhir K. Pandey \footnote{Electronic mail:
sk$_{_{-}}$iuc@rediffmail.com}} \address{UGC-DAE Consortium for
Scientific Research, University Campus, Khandwa Road, Indore -
452001, India} \address{School of Engineering, Indian Institute of
Technology Mandi, Mandi - 175001, India}

\date{\today}

\begin{abstract}
In the present work we study the effect of nitrogen (N) and fluorine
(F) doping in the electronic properties of ZrO$_2$ by using
\emph{ab initio} electronic structure calculations. Our calculations
show the importance of on-site Coulomb correlation in estimating the
correct band gap of ZrO$_2$. The N and F doping provide hole and
electron type impurity states in the band gap closer to the top of the
valance band and bottom of the conduction band, respectively. The formation
of such impurity states may be exploited in fabricating a $p$-$n$
junction expected to be useful in making an ultraviolet
light-emitting diode.

\end{abstract}

\maketitle

\section{Introduction}
The doping of N in place of oxygen (O) in the technologically
important oxide semiconductors has attracted a great deal of
attention from last couple of years
\cite{nakano,batzill,cao,wu,pandeyTiO2,slipukhina,elfimov,drera}.
Most of these activities is directed towards the creation of
magnetism by non-magnetic doping in the conventional dilute magnetic
semiconducting materials like TiO$_2$, ZnO, MgO, \emph{etc.}
 \cite{pandeyTiO2,slipukhina,elfimov,drera}. It has been shown that
the doping N at O sites is equivalent to creating holes in these
systems. Recently, we have shown that F doping in TiO$_2$ is
equivalent to electron doping in it \cite{pandeyTiO2}. Thus one
expects that N- and F-doped oxide semiconductors may be useful in
fabricating many technologically important materials where $p$- and
$n$-type conductions are to be exploited.

Here we explore such possibility in a technologically important wide
band gap semiconductor viz. ZrO$_2$ \cite{gutowski,eichler,zheng}.
At room temperature it shows monoclinic structure and it converts
into tetragonal and cubic structures with increasing temperature.
The compound is a direct band gap semiconductor and its band gap
reported to vary from 5 to 7 eV depending upon the structural phases
and experimental methods used in extracting it
\cite{french,ikarashi,jiang}.  It is a crucial refractory material
used in insulation, abrasive, enamel, \emph{etc.} It is also an
important high k-dielectric compound predicted to be used as a gate
dielectric material \cite{puthen,zhao}.

In this work we study the modification of electronic structure of
monoclinic ZrO$_2$ when N and F are doped at O sites by using
supercell method of \emph{ab initio} electronic structure
calculations. The on-site Coulomb interaction is found to be
important in getting the correct band of the compound. The N- and
F-doped TiO$_2$ are found to be $p$- and $n$-type semiconductors,
respectively. It is proposed that these semiconductors may be useful
in fabricating a $p$-$n$ junction which can be used as an efficient
ultraviolet (UV) light-emitting diode (LED).

\section{Computational details}
The spin unpolarized electronic structure calculations of
ZrO$_{2-x}$N$_x$ and ZrO$_{2-x}$F$_x$ (x=0, 1/32, 1/16 and 1/8)
compounds have been carried out by using {\it state-of-the-art}
full-potential linearized augmented plane wave (FP-LAPW) method
\cite{elk}. The lattice parameters used in the calculations are
taken from the literature \cite{mincryst,teufer} which are
corresponding to monoclinic and tetragonal structure. Here one
should keep in mind that on general ground lattice parameters are
expected to change with doping. However, on comparing the ionic
radii of N and F with O one expects only small change in the lattice
parameters which have little effect on the electronic properties of
the compounds studied in present work. Moreover, any such attempts
of computing lattice parameters using Elk code is very time
consuming as one needs to carry out different calculations by
varying four parameters viz. a, b, c and angle beta. The geometry
optimization for ZrO$_2$ was carried out until the total magnitude
of force at each atomic site comes out to be less than 13 meV/Bohr.
The muffin-tin sphere radii were chosen to be 2 Bohr for the Ti atom
and 1.6 Bohr for the N, O, and F atoms. The 48, 24 and 12 atoms per
cell were considered for calculating the self-consistent charge
densities of x=1/32, 1/16 and 1/8 compounds, respectively.

For the exchange correlation functional, we have considered recently
developed generalized gradient approximation (GGA) form of Perdew
{\em et al.} \cite{perdew}. Here it is important to note that local
density approximation (LDA) and GGA based results often
underestimates the band gap of semiconductors and insulators
\cite{zunger}. In order to find the correct band gap one can use
various improved methods like GW approximation, DFT+$U$ and hybrid
functional. These methods are also found to be useful in studying
the electronic structure of many doped semiconductors
\cite{stroppa,sicolo,virot}. In the DFT+$U$ method one needs the
value of $U$ for a particular set of orbitals, whereas in the other
two methods such restriction is relaxed. However, among these
methods DFT+$U$ is simple and computationally less demanding. Thus
we use DFT+$U$ method to study the band gap problem of ZrO$_2$. The
effect of on-site Coulomb interaction among O 2$p$ and Zr 4$d$
electrons are considered within GGA+$U$ formulation of the DFT
\cite{gga+u}. The value of $U$ is varied from 1 to 4 eV. The
self-consistent solutions were obtained by considering 27 k-points
in the irreducible part of the Brillouin zone. The self-consistency
was achieved by demanding the convergence of the total energy to be
smaller than 10$^{-4}$ Hartree/cell.

\section{Results and discussions}
As mentioned above, the ground state structure of ZrO$_2$ is
experimentally found to be monoclinic (space group \emph{P21/c}) and
it shows structural transition from monoclinic to tetragonal (space
group \emph{P42/nmc}) above room temperature. The total energy
calculations of ZrO$_2$ in monoclinic and tetragonal phases also
suggest that the monoclinic structure is a true ground state as its
energy is found to be $\sim$170 meV less than that of tetragonal
structure. In the monoclinic phase each unit cell contains one kind
of Zr atoms and two kinds of O atoms (O1 and O2) occupying the 4$e$
Wyckoff positions. Therefore, each unit cell contains 4 Zr atoms and
8 O atoms and every Zr atom is surrounded by 7 O atoms. The atomic
positions of Zr, O1 and O2 atoms after geometry optimization are
found to be ( 0.2743, 0.0433, 0.2091), (0.0648, 0.3248,0.3512) and
(0.4493, 0.7558, 0.4676), respectively, which are slightly different
from the experimentally obtained values of (0.2758, 0.0404, 0.2089),
(0.069, 0.342, 0.345) and (0.451, 0.758, 0.479) for Zr, O1 and O2
atoms, respectively \cite{mincryst}.

The total density of states (TDOS) and partial density of states
(PDOS) of ZrO$_2$ compound are plotted in Fig. 1.  The insulating
ground state for ZrO$_2$ compound is clearly evident from Fig. 1(a)
where one can see a large band gap of 4 eV. The valance band (VB)
and conduction band (CB) are predominantly consisting of O 2$p$ and
Zr 4$d$ states, respectively as seen from Fig. 1(b). One can also
observe small contributions of Zr 4$d$ and O 2$p$ PDOS in VB and CB,
respectively. Under pure ionic model for the compound one would not
have expected the presence of Zr 4$d$ and O 2$p$ states in the VB
and CB. This behaviour may be considered as a signature for
breakdown of pure ionic character of Zr-O bonds.  Although our
calculated band gap of 4 eV is $\sim$1 eV less than the experimental
one \cite{ikarashi}, however it provides fairly better estimate of
the band gap in comparison to earlier calculated LDA based gaps of
3.35 eV \cite{datta} and 3.58 eV \cite{jiang}. Here one should keep
in mind that the GGA is a better approximation than LDA and use of
GGA is expected to improve the gap. Moreover, FP-LAPW method used in
the present work was found to provide better estimate of the gap in
comparison to other methods \cite{ong}. Thus the use of GGA within
FP-LAPW method could be the possible reason for better estimate of
band gap in the present work.

The correct band gap of ZrO$_2$ are obtained by carrying out GGA+$U$
calculations where we consider different values of $U$ for Zr 4$d$
and O 2$p$ electrons. The use of $U$ for these two orbitals can be
justified by looking at the PDOS shown in Fig. 1(b) where we have
seen the presence of O 2$p$ and Zr 4$d$ states in the valence band
as well as in the conduction band. The $U$ dependent band gap of the
compound is shown in the inset of Fig. 1(a). It is evident from the
inset that the band gap monotonically increases with increasing $U$
and $U$$\approx$3.5 eV provides a correct band gap of 5 eV. This
value of $U$ is a reasonable estimate for 4$d$ and 2$p$ electrons
and closer to those values of $U$ which are found to provide fairly
good electronic and magnetic properties of MgO and Sr$_3$NiRhO$_6$
(3$d$ and 4$d$ electrons system) compounds
\cite{slipukhina,pandeySNRO}.

The energy of monoclinic Zr$_4$O$_7$N is found to be $\sim$448 meV
less than that of tetragonal Zr$_4$O$_7$N. On comparing this energy
difference with that of ZrO$_2$ one can say that the N-doping is
giving more stability to the monoclinic structure. This suggests
that the monoclinic structure is a true ground state structure for
the N-doped compounds studied here. N doping at O site changes the
electronic structure of the compound drastically and N-doped ZrO$_2$
becomes metallic as evident from Fig. 2(a)-1(c). The N 2$p$ states
mainly contributes around the Fermi level ($\epsilon_F$) and one can
see three peak structures corresponding to three crystal-field split
$p$ orbitals. The energy distribution of O 2$p$ and Zr 4$d$ states
remains almost similar to that of pure ZrO$_2$. There is a large
density of unoccupied N 2$p$ states just above the $\epsilon_F$.
This indicates that N doping is equivalent to creating holes in the
system and unoccupied N 2$p$ states form an unoccupied impurity
band. The width of the unoccupied impurity band is increasing with
increase in N content and found to be about 0.11, 0.18 and 0.19 eV
for x=1/32, 1/16 and 1/8 compounds, respectively.

The total energy calculations of Zr$_4$O$_7$F in monoclinic and
tetragonal structures also suggest that the monoclinic phase is a
true ground state as its energy is found to be $\sim$1634 meV less
than that of tetragonal structure. On comparing this energy
difference with that of ZrO$_2$ and Zr$_4$O$_7$N one can say that
the F-doping is providing better stability to the monoclinic
structure, which is a true ground state structure for all the
concentrations studied here. The TDOS of F-doped compounds are shown
in Fig. 3. Contrary to the N-doped compounds the $\epsilon_F$ of the
F-doped compounds pinned at the CB which consists of Zr 4$d$ states.
As opposed to N 2$p$ states crystal-field split three F 2$p$ bands
contribute deep inside the VB and below 10 eV of the $\epsilon_F$.
One may get surprised by this result as the simple picture of
impurity doping in a semiconductor does not appear to hold here. In
this picture replacing some of O by F is equivalent to doping
electrons in the system and one would have expected that an extra F
2$p$ electron should contribute near the bottom of the CB as an
occupied impurity band. However, Zr 4$d$ electrons are found to
contribute in the occupied impurity band. Here it is important to
note that the simple picture of impurity doping in the semiconductor
is historically based on the studies of covalent semiconductors
\cite{ashcroft}. The present system under study is an ionic
semiconductor. In the case of ZrO$_2$ charge neutrality condition
requires that two Zr 4$d$ electrons should be transferred to the O
atoms. However, replacing O by F requires only one Zr 4$d$ electron
to be transferred to F and remaining one Zr 4$d$ electron is
expected to contribute in the CB in accordance with the calculated
result. The width of the occupied impurity band is increasing with
increase in F content and found to be about 0.12, 0.14 and 0.24 eV
for x=1/32, 1/16 and 1/8 compounds, respectively.

The present work clearly shows the $p$- and $n$-type doping when O
is replaced by N and F, respectively and bandwidth of hole and
electron like impurity bands can be tuned by varying the doping
concentration. Thus N- and F-doped ZrO$_2$ may be used to fabricate
a $p$-$n$ junction. Since the ZrO$_2$ is a direct band gap
semiconductor and its band gap lies in the UV range, therefore, such
$p$-$n$ junction is expected to be used in manufacturing the UV-LED.
It is important to note that the light extraction in LEDs is a major
research area where people are trying to get more light so that it
may be useful for various purposes. Reducing the heating effect of
LEDs is another important area of research. Both issues are related
to the higher refractive indices of the materials conventionally
used in manufacturing the LEDs. Higher the refractive index of a
material separating air lower will be the critical angle of total
internal reflection and hence more light will get back to the
material (less will pass through the air) itself resulting in more
heating of it. The refractive indices of most of the materials used
in manufacturing LEDs are more than 3 \cite{ritable}. However, the
refractive index of ZrO$_2$ is $\sim$2.6 at 5 eV and expected to
provide better light extraction capability in the UV range.
Recently, it has been shown that UV radiation of $\sim$265 nm (i.e.
$\sim$4.7 eV) acts as an effective disinfectant \cite{mori}. The
bandwidths of the unoccupied and occupied impurity bands are found
to be about 0.12 eV each for x=1/32 N- and F-doped compounds.
Further, these band gaps can further be increased by increasing the
doping concentrations. Thus, one expects that radiation emitted from
LED manufactured by $p$-$n$ junction of N- and F-doped ZrO$_2$  can
be used as an effective sterilizing agent. At this juncture it is
important to note that the N-doped ZrO$_2$ has already been
synthesized \cite{liu} and comparison of the ionic radii of N and F
with respect to O also suggests that the synthesis of F-doped
ZrO$_2$ may not be a problem. In the light of these observations it
is tempting to suggest that the N- and F-doped ZrO$_2$ are important
$p$- and $n$-type semiconductors, respectively, that are expected to
be used in making many semiconducting devices.

\section{Conclusions}

The effect of N and F doping on the electronic properties of ZrO$_2$
has been investigated by using \emph{ab initio} electronic structure
calculations. In order to find the correct band gap one needs to
consider the on-site Coulomb interactions among the Zr 3$d$ and O
2$p$ electrons. Replacing O by N and F is found to create hole and
electron like impurity states in the band gap closer to the top of
the valance band and bottom of the conduction band, respectively. It
is proposed that such electron and hole doped semiconductors may be
used in fabricating a $p$-$n$ junction. This $p$-$n$ junction is
expected to work as an efficient UV-LED which may be used as a
disinfectant.




\section{Figure Captions:}

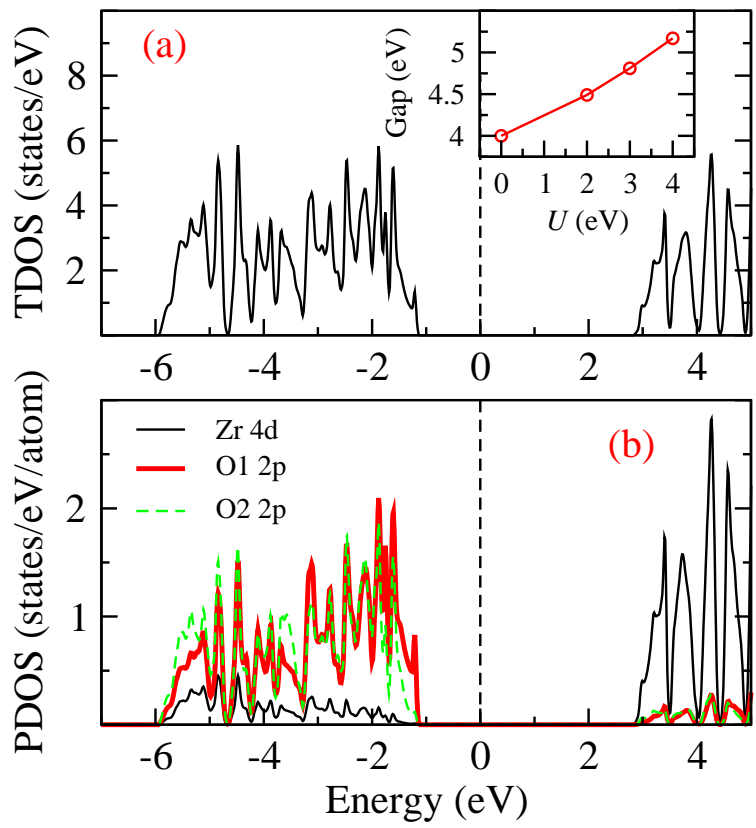
Figure 1: The total density of states (TDOS) and partial density of
states (PDOS) of ZrO$_2$ are shown in (a) and (b). Inset of (a)
shows the evolution of band gap of ZrO$_2$ with increasing the
strength of on-site Coulomb interaction ($U$).

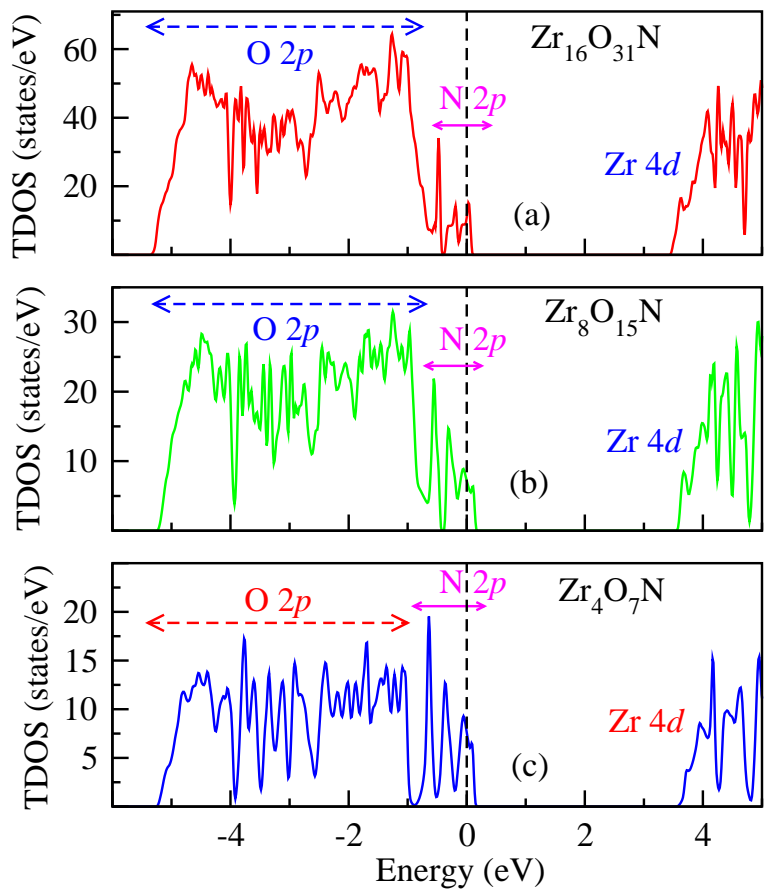
Figure 2: (Color online) Total density of states (TDOS) of
ZrO$_{2-x}$N$_x$ for x=1/32, 1/16, and x=1/8 are shown in (a),
(b), and (c), respectively.

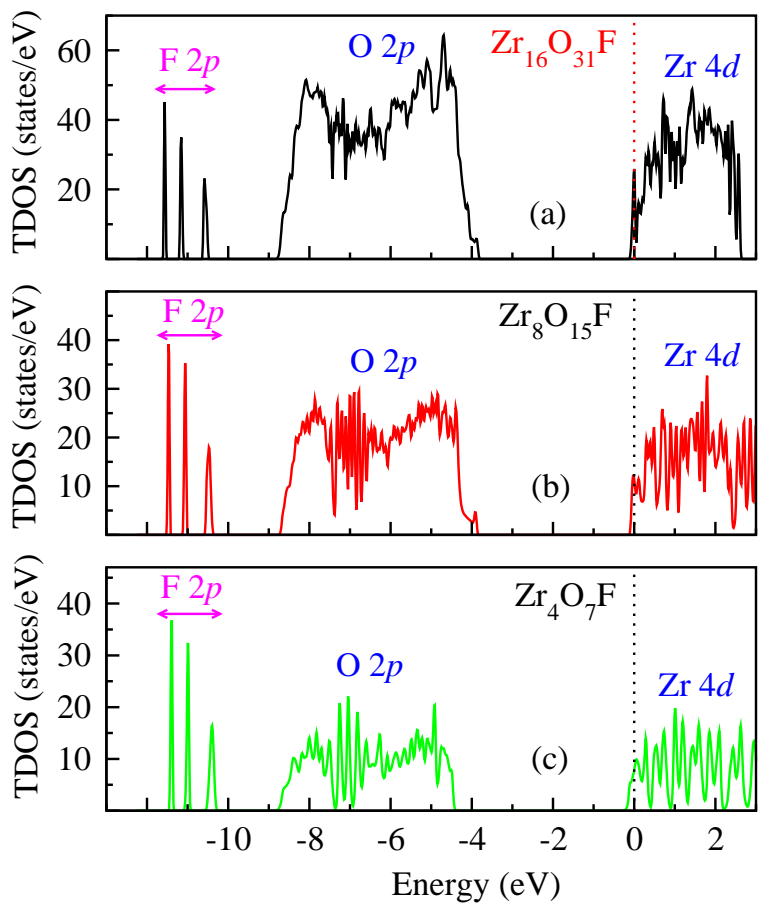
Figure 3: (Color online) Total density of states (TDOS) of
ZrO$_{2-x}$F$_x$ for x=1/32, 1/16, and x=1/8 are shown in (a), (b),
and (c), respectively.





\begin{thebibliography}{99}

\bibitem{nakano} Y. Nakano, T. Morikawa, T. Ohwaki, and Y. Taga,
Appl. Phys. Lett. {\bf 86}, 132104 (2005).

\bibitem{batzill} M. Batzill, E.H. Morales, and U. Diebold,
Phys. Rev. Lett. {\bf 96}, 026103 (2006).

\bibitem{cao} Y. Cao, L. Miao, S. Tanemura, M. Tanemura, Y. Kuno,
and Y. Hayashi, Appl. Phys. Lett. {\bf 88}, 251116 (2006).

\bibitem{wu} H. Wu, A. Stroppa, S. Sakong, S. Picozzi,M. Scheffler,
 and P. Kratzer, Phys. Rev. Lett. {\bf 105}, 267203 (2010).

\bibitem{pandeyTiO2} S.K. Pandey and R.J. Choudhary,
J. Phys.: Condens. Matter {\bf 23}, 276005 (2011).

\bibitem{slipukhina} I. Slipukhina, Ph. Mavropoulos, S. Blugel,
and M. Lezaic, Phys. Rev. Lett. {\bf 107}, 137203 (2011).

\bibitem{elfimov} I.S. Elfimov, A. Rusydi, S.I. Csiszar, Z. Hu, H.H. Hsieh,
H.-J. Lin, C.T. Chen, R. Liang, and G.A. Sawatzky, Phys. Rev. Lett. {\bf 98}, 137202 (2007).

\bibitem{drera} G. Drera, M.C. Mozzati, P. Galinetto, Y. Diaz-Fernandez, L. Malavasi,
F. Bondino, M. Malvestuto, and L. Sangaletti, Appl. Phys. Lett. {\bf 97}, 012506 (2010).

\bibitem{gutowski} M. Gutowski, J.E. Jaffe, C.-L. Liu, M. Stoker,
R.I. Hegde, R.S. Rai, and P.J. Tobin, Appl. Phys. Lett. {\bf 80},
1897 (2002).

\bibitem{eichler} A. Eichler and G. Kresse, Phys. Rev. B {\bf 69}, 045402 (2004).

\bibitem{zheng} J.X. Zheng, G. Ceder, T. Maxisch, W.K. Chim, and
W.K. Choi, Phys. Rev. B {\bf 75}, 104112 (2007).

\bibitem{french} R.H. French, S.J. Glass, F. S. Ohuchi, Y.-N. Xu, and W. Y. Ching,
Phys. Rev. B {\bf 49}, 5133 (1994).

\bibitem{ikarashi} N. Ikarashia and K. Manabe, J. Appl. Phys. {\bf 94}, 480 (2003).

\bibitem{jiang} H. Jiang, R.I. Gomez-Abal, P. Rinke, and M. Scheffler,
Phys. Rev. B {\bf 81}, 085119 (2010).

\bibitem{puthen} R. Puthenkovilakam, E.A. Carter, J.P. Chang,
Phys. Rev. B {\bf 69}, 155329 (2004).

\bibitem{zhao} X. Zhao, D. Ceresoli, and D. Vanderbilt,  Phys. Rev. B {\bf 71}, 085107 (2005).

\bibitem{elk}http://elk.sourceforge.net

\bibitem{mincryst} Y.D. Mc Cullough and K.N. Trueblood, Acta Cryst. {\bf 12}, 95 (1959).

\bibitem{teufer} G. Teufer, Acta Cryst. {\bf 15}, 1187 (1962).

\bibitem{perdew} J.P. Perdew, A. Ruzsinszky, G.I. Csonka, O.A. Vydrov,
G.E. Scuseria, L.A. Constantin, X. Zhou, and K. Burke, Phys. Rev.
Lett., {\bf 100}, 136406 (2008).

\bibitem{zunger} A. Zunger, S. Lany, and H. Raebiger, Physics {\bf 3}, 53 (2010).

\bibitem{stroppa} A. Stroppa and G. Kresse, Phys. Rev. B {\bf 79}, 201201(R)(2009).
A. Stroppa, G. Kresse, and A. Continenza, Phys. Rev. B {\bf 83}, 085201 (2011).

\bibitem{sicolo} S. Sicolo, G. Palma, C. Di Valentin, and G. Pacchioni, Phys. Rev. B {\bf 76}, 075121 (2007).

\bibitem{virot} F. Virot, R Hayn, and A. Boukortt, J. Phys.: Condens. Matter {\bf 23}, 025503 (2011).

\bibitem{gga+u} F. Bultmark, F. Cricchio, O. Gr{\aa}n\"{a}s, and L. Nordstr\"{o}m, Phys. Rev. B
{\bf 80}, 035121 (2009).

\bibitem{datta} G. Dutta, K.P.S.S. Hembram, G.M. Rao, and U.V. Waghmare,
Appl. Phys. Lett. {\bf 89}, 202904 (2006).

\bibitem{ong} K.P. Ong, P. Blaha, and P. Wu, Phys. Rev. B {\bf 77}, 073102 (2008).

\bibitem{pandeySNRO} S.K. Pandey and K. Maiti, Phys. Rev. B {\bf 78}, 045120 (2008).

\bibitem{ashcroft} N.W. Ashcroft and N. Mermin, Solid State Physics,
Chapter 28 (Brooks/Cole Cengage Learning, New Delhi, 1976).

\bibitem{ritable} http://refractiveindex.info

\bibitem{mori} M. Mori, A. Hamamoto, A. Takahashi, M. Nakano, N. Wakikawa,
S. Tachibana, T. Ikehara, Y. Nakaya, M. Akutagawa, and Y. Kinouchi,
Med. Bio. Eng. Comput. {\bf 45}, 1237 (2007).

\bibitem{liu} Y. Liu, J. Li, X. Qiu, and C. Burda, J. Photoch. Phtobio. A {\bf 190}, 94 (2007).


\end{thebibliography}
\end{document}